\documentclass[prl,twocolumn,showpacs,floatfix]{revtex4}

\usepackage{graphicx}
\usepackage{amsmath}

\makeatletter
\def\btt#1{\texttt{\@backslashchar#1}}
\DeclareRobustCommand\bblash{\btt{\@backslashchar}}
\makeatother

\begin{document}

\title{Efficient Universal Leakage Elimination for Physical and Encoded Qubits}
\author{L.-A. Wu, M.S. Byrd, and D.A. Lidar}
\affiliation{Chemical Physics Theory Group, University of Toronto, 80
St. George St., Toronto, Ontario M5S 3H6, Canada }

\date{\today}

\begin{abstract}
Decoherence-induced leakage errors can couple a physical or encoded
qubit to other levels, thus potentially damaging the qubit. They can
therefore be very detrimental in quantum computation and require special
attention. Here we present a general method for removing such errors by using
simple decoupling and recoupling pulse sequences. The proposed gates are
experimentally accessible in a variety of promising quantum computing proposals.
\end{abstract}


\maketitle

The unit of quantum information is the qubit: an idealized two-level system
consisting of a pair of orthonormal quantum states. However, this
idealization neglects other levels which are typically present and can mix
with those defining the qubit. Such mixing, the prevention of which is the
subject of this work, is known as \textquotedblleft
leakage\textquotedblright . Leakage may be the result of the application of
logical operations, or induced by system-bath coupling. In the former case,
a rather general solution was proposed in \cite{Tian:00}. Here we are
interested in decoherence-induced leakage. E.g., in the ion-trap QC proposal
the two-level approximation may break down and spontaneous transitions may
leak population out of those levels that represent the qubit in the ion \cite{Cirac:95Plenio:97}. This is part of a more general problem: quantum computation
(QC)\ depends on reliable components and a high degree of isolation from a
noisy environment. When these conditions are satisfied, it is known that it
is possible to stabilize a quantum computer using an encoding of a
\textquotedblleft logical qubit\textquotedblright\ into several physical
qubits. Methods which profitably exploit such an encoding are, e.g.,
(closed-loop) quantum error correcting codes (QECC) \cite{Preskill:97a,Knill:98} and (open-loop) decoherence-free subspaces or
subsystems (DFS) \cite{Zanardi:97c,Duan:98,Lidar:PRL98,Knill:99a}. The
logical qubits of these codes can also undergo leakage errors,
which are particularly serious: by mixing states from within the code and
outside the code space, leakage completely invalidates the encoding. A
simple procedure to detect and correct leakage, which can be incorporated
into a fault-tolerant QECC circuit, was given in \cite{Preskill:97a}. This
scheme is, however, not necessarily compatible with all encodings \cite{Kempe:01}. Here we present a \emph{universal}, open-loop solution to
leakage elimination, which makes use of fast and strong \textquotedblleft
bang-bang\textquotedblright\ (BB)\ pulses \cite{Viola:98Vitali:99Zanardi:99d,Viola:99}. We first give a general scheme for protecting qubits (whether encoded or
physical) from leakage errors using an efficient pulse sequence. Then we
illustrate the general result with examples taken from a variety of
promising QC\ proposals. Particularly important is the fact that our scheme
is experimentally feasible in these examples, in the sense that we only make
use of the naturally available interactions.

\textit{Universal leakage elimination operator.---} Here we give a general,
existential argument for eliminating all leakage errors on encoded or
physical qubits. Suppose that $n$ two-level systems (e.g., electron spins in
quantum dots \cite{Loss:98Levy:01a}) are used to encode one logical qubit, or that a $N$-level
Hilbert space $\mathcal{H}_{N}$ supports a two-dimensional physical qubit
subspace (e.g., hyperfine energy levels of an ion \cite{Cirac:95Plenio:97}). Let us arrange the basis
vectors $\{\left| n\right\rangle \}_{n=0}^{N-1}$ of $\mathcal{H}_{K}$ so
that $\left| 0\right\rangle $ and $\left| 1\right\rangle $ represent the
(physical or encoded) qubit states ($N=2^{n}$ for the encoded case). We
refer to this as the ``ordered basis''. In this basis we can classify all
system operators as follows:
\begin{equation}
E=\left( 
\begin{array}{cc}
B & 0 \\ 
0 & 0
\end{array}
\right) \quad E^{\bot }=\left( 
\begin{array}{cc}
0 & 0 \\ 
0 & C
\end{array}
\right) \quad L=\left( 
\begin{array}{cc}
0 & D \\ 
F & 0
\end{array}
\right) ,  \label{eq:EEL}
\end{equation}
where $B$ and $C$ are $2\times 2$ and $(N-2)\times (N-2)$ blocks
respectively, and $D,F$ are $2\times (N-2)$, $(N-2)\times 2$ blocks.
Operators of type $E$ represent logical operations, i.e., they act entirely
within the qubit subspace. $E^{\bot }$ operators, on the other hand, have no
effect on the qubit as they act entirely outside the qubit subspace.
Finally, $L$ represents the leakage operators. The total system-bath
Hamiltonian can be written as $H_{SB}=H_{E}+H_{E^{\bot }}+H_{L},$ where $
H_{E}$ ($H_{E^{\bot }},H_{L}$) is a linear combination of elements of the
set $E$ ($E^{\bot },L$), tensored with bath operators. Now consider
\begin{equation}
R_{L}=e^{i \phi}\left( 
\begin{array}{cc}
-I & 0 \\ 
0 & I
\end{array}
\right) ,  \label{eq:R_Lmat}
\end{equation}
where the blocks have the same dimensions as in Eq.~(\ref{eq:EEL}) and
$\phi$ is an overall phase. This
operator satisfies $\{R_{L},L\}=0$, while $[R_{L},E]=[R_{L},E^{\bot }]=0$.
Using a BB parity-kick sequence \cite{Viola:98Vitali:99Zanardi:99d} it follows that $
R_{L}$ is a \emph{leakage-elimination operator} (LEO): 
\begin{equation}
\lim_{n\rightarrow \infty }(e^{-iH_{SB}t/n}R_{L}^{\dagger
}e^{-iH_{SB}t/n}R_{L})^{n}=e^{-iH_{E}t}e^{-iH_{E^{\bot }}t}
\label{eq:BB}
\end{equation}
In practice one takes $n=1$ and makes the time $t$ very small compared to
the bath correlation time \cite{Viola:98Vitali:99Zanardi:99d}. Eq.~(\ref{eq:BB}) then
holds to order $t^{2}$, and implies that one intersperses periods of free
evolution for time $t$ with $R_{L},R_{L}^{\dagger }$ pulses which are so
strong that $H_{SB}$ is negligible during these BB pulses.
The term $e^{-iH_{E^{\bot }}t}$ in Eq.~(\ref{eq:BB}) has no effect on the qubit
subspace. The term $e^{-iH_{E}t}$ may result in logical errors, which will
have to be treated by other methods, e.g., concatenation with a QECC
\cite{Preskill:97a,Knill:98,Lidar:PRL99,Lidar:00b}, or additional BB pulses
\cite{Viola:99,ByrdLidar:01a}. Note that since $R_L$ commutes with the
logical operations, they can be performed at the same time, i.e., {\em
our leakage elimination procedure is fully compatible with universal
QC}. We now give a procedure for generating LEOs from a 
controllable system Hamiltonian $H_{S}$ acting for a time $\tau $, i.e., $
R_{L}=\exp (-iH_{S}\tau )$. From Eq.~(\ref{eq:R_Lmat}) it follows that $
H_{S}$ must act as a projection operator $P$ onto the qubit subspace.
Furthermore, $\tau $ must be chosen so that $R_{L}$ acts as $-I$ in the
qubit subspace. A general choice is 
\begin{equation}
R_{L}^{\mathrm{E}(1)}=\exp \left( \pm i\pi \widehat{n}\cdot \vec{
\sigma }P\right)   \label{eq:R_L^E}
\end{equation}
where $\vec{\sigma }$ denotes the vector of Pauli matrices, which
we refer to as logical $X,Y,Z$ operations, and $\widehat{n}$ is a real unit
vector. This is a valid LEO\ since $\exp \left( -i\pi \widehat{n}\cdot 
\vec{\sigma }\right) $ expresses a $2\pi $ rotation about the
axis $\widehat{n}$ on the qubit Bloch sphere, upon which the qubit state
acquires a minus sign. A useful example is $\tau =\pi $ and $H_{S}=|0\rangle
\langle 0|+|1\rangle \langle 1|$, which is a projector onto the qubit
subspace and acts as identity there. This example generalizes immediately to 
$d$-dimensional qu$d$its:
\begin{equation}
R_{L}^{\mathrm{id}(d)}=\exp (\pm i\pi \sum_{k=0}^{d-1}|k\rangle
\langle k|).
\label{eq:R_Lid}
\end{equation}

Now let us consider leakage prevention on a code subspace of $K$ logical
qubits, each supported by $n$ physical qubits. In analogy to $R_{L}^{\mathrm{
E}(1)}$ we can construct a general LEO as follows. Let $S_{i}$ be a
single-qubit logical (unitary) operation on the $i^{\mathrm{th}}$ (encoded
or physical)\ qubit, and let $P_{i}$ be a projection on the code subspace of
that qubit. Then 
\begin{equation}
R_{L}^{\mathrm{E}(K)}=\exp (\pm i\pi \otimes _{i=1}^{K}S_{i}P_{i})
\label{eq:R_L^EK}
\end{equation}
is a valid LEO.

\textit{Proof}: We can always rotate the Bloch sphere of a
qubit so that each $S_{i}$ is independently transformed into $Z_{i}$:\ $
S_{i}=U_{i}Z_{i}U_{i}^{\dagger }$ (where $U_{i}$ is an appropriate
single-qubit unitary); $\otimes _{i=1}^{K}Z_{i}$ is a diagonal matrix of $
\pm 1$, so $\exp (-i\pi \otimes _{i=1}^{K}Z_{i})=-I$. Thus $R_{L}^{\mathrm{E}
(K)}=\exp (-i\pi \otimes _{i=1}^{K}U_{i}Z_{i}U_{i}^{\dagger }P_{i})=(\otimes
_{i=1}^{K}U_{i})\exp (-i\pi \otimes _{i=1}^{K}Z_{i}P_{i})(\otimes
_{i=1}^{K}U_{i}^{\dagger })=-I$ on the code subspace, and (because of
the $P_i$) $=I$ on the orthogonal complement. QED.

Sometimes we shall be able to construct
single-qubit logical operators which are automatically projectors on that
qubit's subspace. We refer to such operators as ``canonical''. We are now
ready to apply these considerations to a number of promising QC\ proposals.

\textit{Example 1.---} As a simple first example, consider physical qubits
(without encoding), such as electrons on liquid helium \cite{Platzman:99},
or an electron-spin qubit in quantum dots \cite{Loss:98Levy:01a,Imamoglu:99Pazy:01}, or a nuclear-spin qubit in donor
atoms in silicon \cite{Kane:98Vrijen:00}. In those cases, a potential well
at each site traps one fermion. Usually, the ground and first excited state
are taken as a qubit for a given site: $|k\rangle =c_{k}^{\dagger }|\mathrm{vac}
\rangle $, where $c_{k}^{\dagger }$ is a fermionic creation operator for
level $k=0,1$. Let $n_{k}=c_{k}^{\dagger }c_{k}$ be the fermion number
operator. The logical operations for this qubit are
$E=\{X=c_{0}^{\dagger}c_{1}+c_{0}^{\dagger }c_{1},Y=i(c_{1}^{\dagger}c_{0}-c_{0}^{\dagger}c_{1}),Z=n_{0}-n_{1}\}$ whose elements satisfy
$su(2)$ commutation
relations. In this case, a general linear Hamiltonian which includes hopping
terms, $H_{SB}=\sum_{k,l=0}^{N-1}a_{kl}c_{k}^{\dagger }c_{l}$, where $a_{kl}$
includes parameters and bath operators, and $k,l$ denote all electron
states, can leak the qubit states $k=0,1$ into any of the other states.
Using parity-kicks, we can eliminate this leakage in terms of the LEO
[recall Eq.~(\ref{eq:R_Lid})]: $
R_{L}^{\mathrm{id}(1)}=\exp [\pm i\pi (n_{0}+n_{1})]$. This LEO\ is
implemented simply by controlling on-site energies.

\begin{figure}
\includegraphics[height=13cm,angle=0]{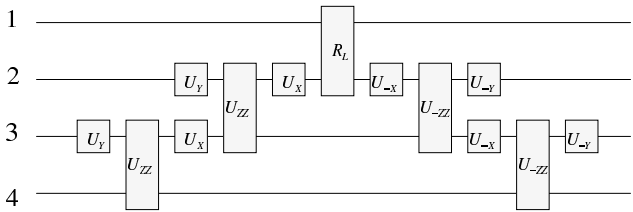}
\vspace{-10cm}
\caption{\label{fig1}
Quantum circuit of the LEO for $4$ physical (encoded) qubits in Example 1
(3). $U_{A}=e^{i\frac{\pi }{4}A}$, where $A$ stands for $
X_{i},Y_{i},Z_{i},Z_{i}Z_{j}$ ($\overline{X}_{i},\overline{Y}_{i},\overline{Z
}_{i},\overline{Z}_{i}\overline{Z}_{j}$) for Example 1 (3).
}
\end{figure}

Let us now generalize this to $K$ qubits. The states of the $i^{\mathrm{th}}$
qubit are $|k\rangle _{i}=c_{k}^{\dagger }(i)|\mathrm{vac}\rangle $. States
outside of the code subspace contain at least one creation operator $
c_{k}^{\dagger }(i)$ with $k\geq 2.$ A logical $Z$ operator on the $i^{
\mathrm{th}}$ qubit is $Z_{i}=n_{0}(i)-n_{1}(i)$, which is canonical. It
follows from Eq.~(\ref{eq:R_L^EK}) that an LEO is: 
\begin{equation}
(R_{L}^{\mathrm{E}(K)})_{\mathrm{ferm}}=\exp (\pm i\pi (Z_{1}Z_{2}\cdots
Z_{K})).
\label{eq:LEO-ferm}
\end{equation}
The term $\otimes _{i=1}^{K}Z_{i}$ involves a many-body interaction which is
not naturally available. However, it can be constructed from available
interactions as follows: Let us assume that the interaction between
neighboring sites $i,j$ contains a controllable $Z_{i}Z_{j}$ term (in
reality such control may have to be obtained indirectly, e.g., by controlling
an $X_{i}X_{j}+Y_{i}Y_{j}$ term, as shown in \cite{LidarWu:01}, and as
discussed in more detail below). We note the following useful ``conjugation
by $\pi /4$''\ formula \cite{Lidar:00b}: 
\begin{equation}
T_{A}\circ e^{i\theta B}\equiv e^{i\frac{\pi }{4}A}e^{i\theta B}e^{-i\frac{
\pi }{4}A}=e^{i\theta (iAB)},
\end{equation}
which holds provided $\{A,B\}=0$ and $A^{2}=B^{2}=I$. Using this we can
efficiently generate long-range interactions by alternately switching
interactions $A,B$ on/off. E.g., 
\begin{equation}
T_{Y_{2}}\circ \left[ T_{Z_{2}Z_{3}}\circ \left( T_{X_{2}}\circ e^{i\theta
Z_{1}Z_{2}}\right) \right] =e^{i\theta Z_{1}Z_{2}Z_{3}},  \label{eq:ZZZ}
\end{equation}
which can, in turn, be used to generate $e^{i\theta Z_{1}Z_{2}Z_{3}Z_{4}}$, etc. Using
this recursive process, the implementation of the LEO $(R_{L}^{\mathrm{E}
(K)})_{\mathrm{ferm}}$ takes $O(K)$ steps. Fig.~\ref{fig1} shows a
circuit for the $4$-qubit case.

Finally, we note that we can also treat bosonic systems, such as the linear
optical QC proposal \cite{Knill:00}. In this case, a qubit is encoded into
two modes. The first qubit has states $|0\rangle _{1}=b_{1}^{\dagger }|
\mathrm{vac}\rangle $ and $|1\rangle _{1}=b_{2}^{\dagger }|\mathrm{vac}
\rangle $, and the second qubit is $|0\rangle _{2}=b_{3}^{\dagger }|\mathrm{
vac}\rangle $ and $|1\rangle _{2}=b_{4}^{\dagger }|\mathrm{vac}\rangle $,
where $b_{i}^{\dagger }$ are bosonic creation operators. Encoded two-qubit
states are $|00\rangle =b_{1}^{\dagger }b_{3}^{\dagger }|\mathrm{vac}\rangle
,$ $|01\rangle =b_{1}^{\dagger }b_{4}^{\dagger }|\mathrm{vac}\rangle ,\
|10\rangle =b_{2}^{\dagger }b_{3}^{\dagger }|\mathrm{vac}\rangle $ and $
|11\rangle =b_{2}^{\dagger }b_{4}^{\dagger }|\mathrm{vac}\rangle .$ But the
linear optical Hamiltonian $H=\sum_{k,l=1}^{4}a_{kl}b_{k}^{\dagger }b_{l}$,
contains beam-splitter terms like $b_{1}^{\dagger }b_{3}$ and $
b_{2}^{\dagger }b_{3}$, which can cause leakage into states such as $
b_{1}^{\dagger }b_{2}^{\dagger }|\mathrm{vac}\rangle $ or $b_{1}^{\dagger 2}|
\mathrm{vac}\rangle $. By using the LEO $R_{L}^{\mathrm{id}(1)}=\exp \left[
\pm i\pi (b_{1}^{\dagger }b_{1}+b_{2}^{\dagger }b_{2})\right] $, we can
eliminate the leakage terms. This LEO can be implemented simply using a
linear optical phaseshifter. However, generalizing this LEO\ to multiple
encoded qubits requires a photon-photon interaction, which is not readily
available.

\textit{Example 3}.--- A substantial number of promising solid-state QC\
proposals, e.g. \cite{Platzman:99,Loss:98Levy:01a,Imamoglu:99Pazy:01,Kane:98Vrijen:00,Mozyrsky:01}, are governed by effective isotropic and anisotropic exchange
interactions, which quite generally, can be written as 
\begin{equation}
H_{\mathrm{ex}}=
\sum_{i<j}J_{ij}^{x}X_{i}X_{j}+J_{ij}^{y}Y_{i}Y_{j}+J_{ij}^{z}Z_{i}Z_{j}.
\label{eq:Hex}
\end{equation}
The encoding $\left| 0\right\rangle _{L}=|01\rangle $, $\left|
1\right\rangle _{L}=|10\rangle $ (using two physical qubits per logical
qubit) is highly compatible with $H_{\mathrm{ex}}$, in the sense that
universal QC can be performed by controlling the \emph{single} parameter $
J_{ij}^{x}$ in the Heisenberg ($J_{ij}^{x}=J_{ij}^{y}=J_{ij}^{z}$), XXZ ($
J_{ij}^{x}=\pm J_{ij}^{y}\neq J_{ij}^{z}$), and XY ($J_{ij}^{x}=J_{ij}^{y}$, 
$J_{ij}^{z}=0$) instances of $H_{\mathrm{ex}}$, provided there is a Zeeman
splitting that distinguishes single-qubit $Z_{i}$ terms. This is done using
the ``encoded selective recoupling''\ method \cite{LidarWu:01}.
Furthermore, the $\{|01\rangle ,|10\rangle \}$ encoding is a DFS for
collective dephasing (where the bath couples only to system $Z_{2i-1}+Z_{2i}$
operators) \cite{Duan:98,Lidar:PRL99,Kempe:00}. A set of logical operations
on the code is $E=\{\overline{X}
_{1}=(X_{1}X_{2}+Y_{1}Y_{2})/2$, $\overline{Y}_{1}=(X_{2}Y_{1}-Y_{2}X_{1})/2$, $\overline{Z}_{1}=(Z_{1}-Z_{2})/2\}$. Only the $\overline{X}_{1}$ term is
assumed to be directly controllable (by manipulation of $J_{12}^{x}$), while
the $\overline{Z}_{1}$ term can be turned on/off using recoupling \cite{LidarWu:01}. The $\overline{Y}_{1}$ term can then be reached in a few
steps: $e^{-i\theta \overline{Y}_{1}}=T_{\overline{X}_{1}}\circ e^{-i\theta 
\overline{Z}_{1}}$. The leakage errors are due to system-bath interactions
where the system terms include any of $X_{i},Y_{j},$ $X_{i}Z_{j}$ and $
Y_{i}Z_{j}$, since as is easily seen, such terms do not preserve the $
\{|01\rangle ,|10\rangle \}$ code subspace. As pointed out first in \cite
{ByrdLidar:01a}, the LEO can be expressed as $(R_{L}^{\mathrm{E}(1)})_{2-
\mathrm{DFS}}=\exp (i\pi \overline{X}_{1})=Z_{1}Z_{2}$, which means that it
is implementable using just the controllable $J_{12}^{x}$ parameter in the
instances of $H_{\mathrm{ex}}$ mentioned above. This form for $(R_{L}^{
\mathrm{E}(1)})_{2-\mathrm{DFS}}$ is an instance of Eq.~(\ref{eq:R_L^E}),
with $\widehat{n}=\widehat{x}$. Note that, in agreement with our
general comments above,  $(R_{L}^{\mathrm{E}(1)})_{2-
\mathrm{DFS}}$ commutes with every element of $E_{2-\mathrm{DFS}}$, meaning
that logical operations can be performed on the encoded subspace \emph{while
eliminating leakage}.

Next we now show how to efficiently eliminate leakage in this case on an 
arbitrary number of encoded qubits. The $m^{\mathrm{th}}$ logical
qubit is encoded as $|0_{L}\rangle _{m}=|0_{2m-1}1_{2m}\rangle $, $
|1_{L}\rangle _{m}=|1_{2m-1}0_{2m}\rangle $. The logical $Z$ operator is $
\overline{Z}_{m}=(Z_{2m-1}-Z_{2m})/2$, and is canonical. It follows from
Eq.~(\ref{eq:R_L^EK})  that a valid LEO is: 
\begin{equation}
(R_{L}^{\mathrm{E}(K)})_{2-\mathrm{DFS}}=\exp (\pm i\pi \overline{Z}_{1}
\overline{Z}_{2}...\overline{Z}_{K}).
\end{equation}
The next question is how to efficiently construct such an operator. The term 
$\overline{Z}_{1}\overline{Z}_{2}=\frac{1}{4}
(Z_{1}Z_{3}+Z_{2}Z_{4}-Z_{2}Z_{3}-Z_{1}Z_{4})$ contains next and second-next
nearest-neighbor interactions. Using ``conjugation by $\pi /4$'', they can
all be generated using only nearest-neighbor interactions in terms of the
relation 
\begin{eqnarray}
e^{i\theta Z_{i}Z_{i+2}} &=&T_{X_{i+1}X_{i+2}}\circ \left(
T_{Y_{i+1}Y_{i+2}}\circ e^{i\theta Z_{i}Z_{i+1}}\right)   \notag \\
&=&T_{X_{i+1}X_{i+2}+Y_{i+1}Y_{i+2}}\circ e^{i\theta Z_{i}Z_{i+1}}
\label{eq:ZiZi+2}
\end{eqnarray}
where, in accordance with \cite{LidarWu:01}, we have only assumed
controllability of the XY\ interaction term $X_{i+1}X_{i+2}+Y_{i+1}Y_{i+2}$.
At this point we can use the recursive construction of Eq.~(\ref{eq:ZZZ})
again, by replacing $X,Y,Z$ there by their encoded counterparts. Doing so
takes $\overline{Z}_{i}\overline{Z}_{i+1}$ to $\overline{Z}_{i}\overline{Z}
_{i+1}\overline{Z}_{i+2}$, etc., and will again \emph{efficiently} construct
the LEO $(R_{L}^{\mathrm{E}(K)})_{2-\mathrm{DFS}}$, i.e., using $O(K)$
steps. An example of this for $4$ encoded qubits is shown in Fig.~\ref{fig1}. It is
interesting to contrast the linear scaling of this leakage elimination
procedure with general error elimination using BB\ pulses. As shown in \cite{Viola:99}, without additional symmetry assumptions restricting the order of
coupling terms in the Hamiltonian Eq.~(\ref{eq:BB}), the BB procedure, if used
to eliminate \emph{all errors}, requires a number of pulses that is
exponential in $N$.

\textit{Example 4.---} Collective decoherence is a system-bath interaction
that obeys full qubit permutation symmetry:\ $H_{SB}^{\mathrm{Coll.Dec.}
}=\sum_{\alpha =x,y,z}\left( \sum_{i}\sigma _{i}^{\alpha }\right) \otimes
B^{\alpha }$, where $\sigma _{i}^{\alpha }$ are the Pauli matrices and $
B^{\alpha }$ are bath operators \cite{Zanardi:97c,Duan:98,Lidar:PRL98}. This
situation can be \emph{created} from an arbitrary linear system-bath
coupling $H_{SB}^{(1)}=\sum_{i}\vec{\sigma}_{i}\cdot \vec{B}_{i}$, where $\vec{\sigma}_{i}=(X_{i},Y_{i},Z_{i})$ and $\vec{B}_{i}$
are bath operators, using a BB symmetrization pulse-sequence that employs 
\emph{only} the Heisenberg exchange interaction \cite{WuLidar:01b}. The
shortest DFS (or ``noiseless subsystem'') encoding that protects a single
logical qubit against collective decoherence uses $3$ physical qubits \cite{Knill:99a}. In \cite{Kempe:00} it was shown that one can perform universal
QC on this DFS, again using only the Heisenberg interaction. To explain the
encoding, note that the Hilbert space of $3$ spin $1/2$'s has total-spin $
\vec{S}=\frac{1}{2}(\vec{\sigma}_{1}+\vec{\sigma}_{2}+\vec{\sigma}
_{3})$ and splits into two $S=1/2$ subspaces (denoted $\lambda =0,1$), and a 
$S=3/2$ subspace. The states can be labeled as $|S,\lambda ,S_{z}\rangle $,
and the DFS qubit is $|0_{L}\rangle =\alpha |1/2,0,1/2\rangle +\beta
|1/2,0,-1/2\rangle $, $|1_{L}\rangle =\alpha |1/2,1,1/2\rangle +\beta
|1/2,1,-1/2\rangle $, $|\alpha |^{2}+|\beta |^{2}=1$, i.e., the encoding is
into the degeneracy of the two $S=1/2$ subspaces \cite{Knill:99a,Kempe:00}.
Collective errors can change the $\alpha ,\beta $ coefficients, but have the
same effect on the $|0_{L}\rangle ,|1_{L}\rangle $ states, which is why this
encoding is a DFS. If, however, we also consider bilinear system-bath
coupling $H_{SB}^{(2)}=\sum_{i<j}\sum_{\alpha ,\beta
=\{x,y,z\}}g_{ij}^{\alpha \beta }\sigma _{i}^{\alpha }\sigma _{j}^{\beta
}\otimes B_{ij}^{\alpha \beta }$, where $g_{ij}^{\alpha \beta }$ is a rank-$2
$ tensor, then the symmetrization procedure of \cite{WuLidar:01b}, that
prepares collective decoherence conditions, will not work. In this case we
must consider the possibility of leakage. The bilinear term $g_{ij}^{\alpha
\beta }\sigma _{i}^{\alpha }\sigma _{j}^{\beta }$ can be decomposed into (i)
a scalar $g\vec{\sigma}_{i}\cdot \vec{\sigma}_{j}$, which has the effect of
logical errors $E$; (ii) two rank-$1$ tensors $\vec{\beta }\cdot $ $(
\vec{\sigma}_{i}\times \vec{\sigma}_{j})$ and $
\left( \vec{\sigma}_{i}\cdot \vec{\gamma }\right) \left( \vec{
\sigma}_{j}\cdot \vec{\gamma }\right) $, which can couple between $S=1/2$
states, and can couple them to $S=3/2$ states. Note that this also
applies to {\em imperfect symmetrization} at the level of a {\em
linear} system-bath Hamiltonian $H_{SB}^{(1)}$. Thus we see that the $S=3/2$
subspace acts as a source for leakage, and that there
is also the possibility of (non-collective) errors [both from (ii)] which do not
have the same effect on the $|0_{L}\rangle ,|1_{L}\rangle $
states. We defer an analysis of the latter ``$S=1/2 \rightarrow 1/2$''
errors to a separate publication \cite{ByrdWuLidar:tbp}, but we note
that they can be suppressed using techniques similar to those we
discuss next.

An open-loop leakage correction circuit for this DFS, that once more uses
only the Heisenberg interaction, was given in \cite{Kempe:01}. There the
DFS qubit was defined to be $|0_{L}\rangle =|1/2,0,1/2\rangle $, $
|1_{L}\rangle =|1/2,1,1/2\rangle $ and transitions to any of the other $6$
states were considered as leakage (this includes errors caused by
collective decoherence, which are normally avoided by a DFS encoding). Here we add another element to this
picture of \emph{the Heisenberg interaction as an enabler of
universal, fault-tolerant QC}, by showing that it can also provide an LEO. The
importance of Heisenberg-only QC, as pointed out in \cite{DiVincenzo:00a},
is in the relative ease of manipulating this interaction in a number of the
most promising solid-state QC proposals \cite{Loss:98Levy:01a,Kane:98Vrijen:00}. Now, as shown in \cite{Kempe:00}, $
\overline{X}=\frac{1}{4\sqrt{3}}(\vec{\sigma}_{1}\cdot \vec{\sigma}_{3}-\vec{
\sigma}_{2}\cdot \vec{\sigma}_{3})$ acts as a logical $X$ on the DFS qubit
defined above, and annihilates the $S=3/2$ states (i.e., it is canonical).
Therefore, using Eq.~(\ref{eq:R_L^E}), $(R_{L}^{\mathrm{E}(1)})_{3-\mathrm{DFS
}}=\exp (\pm i\pi \overline{X})$ is a Heisenberg-only LEO for a single $3$-qubit DFS which eliminates transitions to the $S=3/2$ subspace. An LEO for
the $K$-qubit case is then, from Eq.~(\ref{eq:R_L^EK}): 
\begin{equation}
(R_{L}^{\mathrm{E}(K)})_{3-\mathrm{DFS}}=\exp (\pm i\pi (\overline{X}_{1}
\overline{X}_{2}\cdots \overline{X}_{K})).
\label{eq:R_L3DFS}
\end{equation}
To generate this LEO from available interactions we use a procedure similar
to Eq.~(\ref{eq:ZZZ}). First, note that $T_{\overline{Y}_{2}}\circ T_{
\overline{Z}_{1}}\circ T_{\overline{Z}_{1}\overline{Z}_{2}}\circ e^{i\theta 
\overline{X}_{1}}=e^{i\theta \overline{X}_{1}\overline{X}_{2}}$. Efficient
schemes for generating $\overline{Z}_{i}\overline{Z}_{j}$ were given in \cite{DiVincenzo:00a}, while $\overline{Z}_{i},\overline{Y}_{i}$ are directly
obtainable from the Heisenberg interaction \cite{Kempe:00}. The recursive
construction of $(R_{L}^{\mathrm{E}(K)})_{3-\mathrm{DFS}}$ then proceeds
using $T_{\overline{Y}_{3}}\circ T_{\overline{Z}_{2}}\circ T_{\overline{Z}
_{2}\overline{Z}_{3}}\circ e^{i\theta \overline{X}_{1}\overline{X}
_{2}}=e^{i\theta \overline{X}_{1}\overline{X}_{2}\overline{X}_{3}}$, etc,
which again is a procedure that scales as $O(K)$.

\textit{Conclusions.---} Decoherence-induced leakage from the logical space
of (physical or encoded) qubits is a severe source of errors for quantum
computation. We have shown how to efficiently and universally eliminate such
errors using sequences of \textquotedblleft bang-bang\textquotedblright\
pulses. These pulses can be applied at the same time as logical
operations, so that leakage elimination can be performed in
conjunction with universal quantum computation. Applications to a variety of promising quantum computing proposals
were discussed, and leakage elimination  methods were presented that are directly applicable
using only experimentally available interactions.

\begin{acknowledgments}
The present study was sponsored by the DARPA-QuIST program (managed by
AFOSR under agreement No. F49620-01-1-0468), by PRO, PREA, and the Connaught
Fund (to D.A.L.). We thank Dr. S. Schneider for helpful discussions.
\end{acknowledgments}

\end{document}